\documentstyle[prb,aps,graphicx,multicol]{revtex}
\begin{document}  
\draft
\title{Circuit Effect On The Current-Voltage Characteristics Of Ultrasmall
Tunnel Junctions}
\author{X. H. Wang and K. A. Chao}
\address{Department of Theoretical Physics, Lund University, \\
Helgonav{\"a}gen 5, S-223 62 Lund, Sweden}
\date{PREPRINT}
\maketitle
\begin{abstract}
We have used the method of generating functional in imaginary time to derive
the current-voltage characteristics of a tunnel junction with arbitrary
tunneling conductance, connected in series with an external impedance and a
voltage source. We have shown that via the renormalized charging energy and
the renormalized environment conductance, our nonperturbative expressions of
the total action can be mapped onto the corresponding perturbative formulas.
This provides a straightforward way to go beyond the perturbation theory.
For the impedance being a pure resistance, we have calculated the
conductance for various voltages and temperatures, and the results agree
very well with experiments.
\end{abstract}


\begin{multicols}{2}

\section{Introduction}

By decreasing the capacitance $C$ of a double tunnel junction structure, the 
role of Coulomb charging energy $e^2/2C$ becomes more important. For example, 
if $C$ is of the order 1 $fF$ and the temperature $T$ is lower than 1 $K$, the
elementary Coulomb charging energy $e^2/2C$ dominates the charge transport
through a tunnel junction system, provided that the tunneling resistance
$R_T$ is larger than the resistance quantum $R_K$=$h/e^2$. The technology
nowadays can fabricate tunnel junctions with $C$ less than 1 $fF$, and hence
phenomena related to Coulomb blockade (CB) have been extensively studied in
recent years\cite{alt91,gd92,sz90}.

For a tunnel junction system with $R_T$ much larger than $R_K$=$h/e^2$, the
perturbation theory of CB has been well established, where one performs low
order perturbation expansion in terms of the tunneling 
Hamiltonian\cite{alt91,gd92,sz90}. The so-obtained theoretical results are
in good agreement with experimental observations in multijunction
structures\cite{alt91,gd92,sz90,dev90,gir90}. On the other hand, for a single
tunnel junction, one has to take into account the effect of circuit elements
which are attached to the single junction in order to perform measurements.
If the circuit impedance is much smaller than the quantum resistance $R_K$,
for example, the free space impedance ${\cal Z}_f$$\simeq$377~$\Omega$, the
CB effect can hardly be detected. However, the CB phenomena emerge when a
large impedance is inserted into the circuit in series connection with the
single junction\cite{gd92,dev90,gir90}. In a multijunction system, the effect
of circuit impedance is secondary because there are more than one large
junction resistance. The influence of circuit elements is a specific type of
electromagnetic environment (EME) effect.

In the weak tunneling regime, Devoret {\it et al\,}\cite{dev90} have used
the Fermi's golden  rule to treat the tunneling Hamiltonian, and have
developed a so-called $P(E)$ theory to study the Coulomb charging effect of
a single tunnel junction surrounded by an EME.
Girvin {\it et al\,}\cite{gir90} have used the Green's function technique
to reach the same theory. They studied a tunnel junction connected in series
with an external impedance and a voltage source. The Hamiltonian of the device
consists of four parts
\begin{equation}
H=H_c+H_{\rm el}+H_t+H_{\rm ex} \, . \label{htot}
\end{equation}
When a charge $Q$ is added to the junction with junction capacitance $C$,
the Coulomb charging energy is
\begin{equation}
H_c=Q^2/2C \, . \label{hc}
\end{equation}
The quasi-particle states in the left and the right electrode are labeled
by $\sigma$ for the spin and the transverse motion. The longitudinal wave
vectors in the right electrode are represented by $k$, and in the left
electrode by $q$. In standard quasi-particle notations, the Hamiltonian
$H_{\rm el}$ for the electrodes has the form
\begin{equation} \label{hel}
H_{\rm el}=\sum_{k\sigma} \varepsilon_{k\sigma} 
c^+_{k\sigma} c_{k\sigma} +
\sum_{q\sigma} \varepsilon_{q\sigma} c^+_{q\sigma} c_{q\sigma} \, , 
\end{equation}
Since $\sigma$ is conserved during the tunneling through a junction, we
use $\sigma$ to specify the tunneling channel. For a fixed tunneling
channel, let $t_{kq\sigma}$ be the matrix element for tunneling from the
state $q\sigma$ in the left electrode to the state $k\sigma$ in the right
electrode. Then, the tunneling Hamiltonian is expressed as
\begin{equation} \label{ht}
H_t=\sum_{kq \sigma} [t_{kq\sigma}c^+_{k\sigma} c_{q\sigma}
\exp (-i\varphi) + {\rm H.c.}] \, ,
\end{equation}
where the operator $\varphi$ is conjugate to $Q$ via the commutation
relation  $[\varphi,Q]$=$ie$. We should mention that here we have neglected
the influences of the tunneling time, which is appropriate for metallic
tunnel junctions\cite{naz89,but86,ave94}. Since the energies of tunneling
electrons are near the Fermi energy, $t_{kq\sigma}$ can be well approximated
by a constant value $t$. In the presence of a bias voltage $V$, the external
impedance is simulated by a set of LC
circuits\cite{gd92,dev90,gir90,l1,cal83} as
\begin{equation} \label{hex}
H_{\rm ex} = \sum_{m=1}^{\infty} \left[
\frac{q_m^2}{2C_m} + \frac{(eVt-\varphi-\varphi_m)^2}{2e^2 L_m}
\right] \, ,
\end{equation}
where the operators $\varphi_m$ and $q_n$ obey the commutation relation
$[\varphi_m,q_n]$=$ie\delta_{m,n}$.  

As we mentioned before, in the regime of weak tunneling, perturbation
theory has been applied to the Hamiltonian Eq.~(\ref{htot}) to calculate
the current as a function of the bias voltage\cite{gd92,dev90,gir90}.
Beyond the weak tunneling regime, following the pioneering work of
Ambegaokar {\it et al\,}\cite{aes82}, the nonperturbative approach of path
integral has been used to investigate the statistical properties\cite{weg}
and the transport behavior\cite{l1} in the linear-response regime of tunnel
junctions (also see Ref.~\onlinecite{sz90} for the current-bias case).
Recently, using the path integral approach formulated in imaginary time to
calculate the linear conductance\cite{l1}, we found that the strong
tunneling processes renormalize both the charging energy and the external
impedance. Furthermore, within this imaginary time formalism, we have shown
that at sufficiently high temperatures, the action of the tunnel junction
can be approximated by the action of the corresponding resistance, and so
an explicit expression of the high temperature quantum conductance can be
derived\cite{l1}. Along the line of Feynman's formalism for polaron, the
Coulomb charging effect has been investigated with the Ohmic 
approximation\cite{odi88}, the nonequilibrium Green's function
method\cite{huo97}, and the mean-field theory\cite{jed98}. The mean-field
theory also demonstrates the normalization of the charging energy and the
external impedance, and the so-calculated zero-bias conductance agrees with
experiment as long as $k_BT$ is not much less than $e^2/2C$ and $R_T$ is
not much smaller than $R_K$.

Nonperturbative approach has so far produced concrete results only for the
situation of zero-bias. In this paper we will consider finite bias and so
investigate the full current-voltage (I-V) characteristics of a
voltage-biased tunnel junction connected in series with an external
impedance. The derivation of the path integral representation for the
generating functional in imaginary time will be outlined in Sec.~II, leaving
the complete mathematical manipulations in the Appendix. With the generating
functional we obtain the tunneling current as a function of the bias voltage.
Specific cases suitable for comparing with experiments will be studied in
details in Sec.~III, showing very good agreement with experiments. Finally,
in Sec.~IV we will prove that via the renormalized charging energy and the
renormalized environment conductance, our nonperturbative expressions of
the total action can be mapped onto the well-established perturbative
results. This provides a straightforward way to go beyond the perturbation
theory.

\section{General Formulation of the I-V Characteristics}

To study the physical system described by Eqs.~(\ref{htot})-(\ref{hex}),
one needs to deal with the electron tunneling and the effect of external
impedance. While the EME has been treated nonperturbatively by many authors,
in this paper we will analyze both the tunneling part and the EME part with
a nonperturbative approach. To achieve this goal, we will start from the
generating functional for a tunneling junction connected in series with an
external impedance and a voltage source.

The effect of the bias voltage $V$ can be incorporated into the Hamiltonian 
not explicitly depending on the time variable via a time-dependent unitary 
transformation defined as 
\[
U(t) = \exp{[ieVt\,(\sum_{k\sigma} c^{\dag}_{k\sigma}c_{k\sigma})]} \, .
\]
It transforms the Hamiltonian in Eq.~(\ref{htot}) to
\end{multicols}
\noindent\hrulefill\hspace{0.5\textwidth}
\begin{eqnarray}
{\cal H} &=& {\cal Q}^2/2C + \sum_{k,\sigma}
(\varepsilon_{k\sigma}+eV) c^{\dag}_{k\sigma} c_{k\sigma} +
\sum_{q,\sigma} \varepsilon_{q\sigma} c^{\dag}_{q\sigma} c_{q\sigma}
+ \sum_{kq\sigma}
[t_{kq\sigma}c^{\dag}_{k\sigma}c_{q\sigma}\exp{(i\phi)} +
{\rm H.c.}]
\nonumber \\
&& + \sum_{n=1}^{\infty} \left[ \frac{q_n^2}{2C_n} +
\frac{(\phi-\varphi_n)^2}{2e^2 L_n} \right] \, ,
\end{eqnarray}
\hspace{0.5\textwidth}\hrulefill
\begin{multicols}{2}
%
\noindent
where ${\cal Q}$=$Q$-$CV$ reflects the quantum fluctuations of the charge
on the tunnel junction, and $\phi$=$eVt$-$\varphi$ serves as the phase of
the external impedance. These two new canonical variables satisfy the
commutation relation $[\phi,{\cal Q}]$=$ie$. The corresponding current 
operator has the form
\begin{equation}
I_T=-ie\sum_{kq\sigma}[t_{kq\sigma}c^{\dag}_{k\sigma}
c_{q\sigma} \exp{(i\phi)}-{\rm H.c.}] \, .
\end{equation}

The generating functional in imaginary time is defined as 
\begin{equation}\label{genfu}
Z_V[\eta] = {\rm tr} \left\{ \hat{T}_\tau \exp{\left\{ -\int_{0}^{\beta} 
d \tau [{\cal H}-I_T \eta(\tau)] \right\}} \right\} \, , 
\end{equation}
where $\hat{T}_\tau$ is the time-ordering operator in imaginary time.
Substituting
\end{multicols}
\noindent\hrulefill\hspace{0.5\textwidth}
\begin{eqnarray}\label{htct}
& & {\cal H}-I_T \eta(\tau) = {\cal Q}^2/2C+\sum_{k,\sigma}
(\varepsilon_{k\sigma}+eV) c^{\dag}_{k\sigma} c_{k\sigma} 
+ \sum_{q,\sigma} \varepsilon_{q\sigma} c^{\dag}_{q\sigma} c_{q\sigma}
\nonumber \\
& &+ \sum_{kq\sigma}
\{ t_{kq\sigma} [1+ie\eta(\tau)] c^{\dag}_{k\sigma}c_{q\sigma}\exp{(i\phi)}
+ {\rm H.c.} \} +\sum_{n=1}^{\infty} \left[ \frac{q_n^2}{2C_n} +
\frac{(\phi-\varphi_n)^2}{2e^2 L_n} \right]
\end{eqnarray}
\hspace{0.5\textwidth}\hrulefill
\begin{multicols}{2}
%
\noindent
into Eq.~(\ref{genfu}), we can rewrite $Z_V[\eta]$ as a series expansion in
terms of the tunneling part of the quantity $\cal H$-$I_T\eta(\tau)$. Then, 
$\beta$ is divided into segments of length
$\varepsilon$ $\rightarrow 0$, and eigenstates of $\phi$ and
${\cal Q}$ are inserted between the elements of the Trotter product. Next,
the charge numbers are summed over. Such mathematical manipulations are
performed in the Appendix, leading to the expression
\end{multicols}
\noindent\hrulefill\hspace{0.5\textwidth}
\begin{eqnarray}\label{genfu1}
& & Z_V[\eta]={\cal N}'' 
\sum_{l=0}^{\infty}(-1)^l \int_0^{\beta} d\tau_l \int_0^{\tau_l}
d\tau_{l-1} \cdots \int_0^{\tau_2} d\tau_1 
\prod_{m=1}^{\infty} \int D\varphi_m 
\int d\phi_0 \prod_{i=1}^{M} \int d\phi_i \sum_{k_1^{}, q_1^{},
\sigma_1^{}, \xi_1^{}} \cdots \sum_{k_l^{}, q_l^{}, \sigma_l^{},
\xi_l^{}} 
\nonumber \\
& &\exp \{-\sum_{i=1}^{M} \varepsilon 
[\dot{\phi}_i^2/4E_c 
+C_m \dot{\varphi}_m^2/2e^2+(\varphi_m-\phi_i)^2/2e^2L_m] \} \nonumber \\
& & \times t^l \xi_1 \xi_2 \cdots \xi_l 
[1+ie\xi_l \eta(\tau_l)] \cdots [1+ie\xi_1 \eta(\tau_1)]
\prod_{\sigma} {\rm tr}_{\sigma} e^{-\beta
h_{\sigma}} h_{k_l^{}, q_l^{}, \sigma_l^{}, \xi_l^{}}(\tau_l) \cdots
h_{k_1^{}, q_1^{}, \sigma_1^{}, \xi_1^{}}(\tau_1) \, ,
\end{eqnarray}
\hspace{0.5\textwidth}\hrulefill
\begin{multicols}{2}
%
\noindent
where
\begin{equation}
h_{\sigma}=\sum_{k} (\varepsilon_{k\sigma}+eV) 
c^+_{k\sigma} c_{k\sigma} +
\sum_{q} \varepsilon_{q\sigma} c^+_{q\sigma} c_{q\sigma}
\end{equation}
and 
\begin{equation}
h_{k,q,\sigma ,\xi}(\tau)= e^{\tau h_\sigma}
c^{\xi}_{k\sigma} c^{-\xi}_{q\sigma}
e^{-\tau  h_\sigma} e^{i \xi \phi(\tau)} \, .
\end{equation}
Here we have introduced the notations $c^{\xi}_{k\sigma}$=$c^+_{k\sigma}$
for $\xi$=1, and $c^{\xi}_{k\sigma}$=$c_{k\sigma}$ for $\xi$=-1. The same
notations apply to $c^{\xi}_{q\sigma}$.
  
The trace over the electronic states in Eq.~(\ref{genfu1}) will be
evaluated by first introducing the quantum statistic average of the
products of creation and annihilation operators, and then using the Wick's
theorem. Here we have made use of the fact that the number of tunneling
channels in metallic junctions, $N$ is very large. Therefore, terms of
order 1/$N$ are neglected. Again, these algebraic works are done in the
Appendix, and the path integral representation of the generating functional 
is finally derived as
\end{multicols}
\noindent\hrulefill\hspace{0.5\textwidth}
\begin{eqnarray}\label{genfu2}
& &Z_V[\eta]=\prod_{m=1}^{\infty} \int D\varphi_m \int D\phi 
e^{-\int_0^\beta d \tau [\dot{\phi}^2/4E_c
+C_m \dot{\varphi}_m^2/2e^2+(\phi-\varphi_m)^2/2e^2L_m]}
\nonumber \\
& & \sum_{r=0}^{\infty} \frac{1}{r!} 
\left\{ \int_0^\beta d\tau \int_0^\beta  d\tau' \alpha_t(\tau-\tau') 
e^{[(eV\tau +i\phi (\tau))]} e^{[-(eV\tau' +i\phi (\tau'))]}
[1+i e \eta(\tau)][1-i e \eta(\tau')] \right\}^r \, ,
\end{eqnarray}
\hspace{0.5\textwidth}\hrulefill
\begin{multicols}{2}
%
\noindent
which is Eq.~(\ref{appen2}) in the Appendix.

In the above generating functional, the path integrals over the EME modes
are Gaussian, and thus can be carried out exactly\cite{com} to give
\begin{equation} \label{zvpi}
Z_V[\eta] =
\int D\phi e^{-S_c[\phi]-S_{\rm ex}[\phi]-S_t[\phi,\eta]} \, .
\end{equation}
In the total action 
\begin{equation}\label{sphi}
S[\phi,\eta] \equiv S_c[\phi] + S_{\rm ex}[\phi] + S_t[\phi,\eta] \, ,
\end{equation}
the first term on the right hand of the above equation 
\begin{equation}
S_c[\phi] = \frac{1}{4E_c} \int_0^{\beta}d\tau {\dot{\phi}}^2
\end{equation}
is resulted from the Coulomb charging energy, the second term
\begin{equation}\label{genfu3}
S_{\rm ex}[\phi] =
\frac{1}{2} \int_{0}^{\beta} d\tau \int_{0}^{\beta} d\tau'
\alpha_{\rm ex}(\tau-\tau')  [\phi(\tau)-\phi(\tau')]^2
\end{equation}
is due to the external impedance, and the third term
\begin{eqnarray}\label{genfu4}
& S_t[\phi, \eta] & =
-\int_{0}^{\beta} d\tau \int_{0}^{\beta}  d\tau' \alpha_t(\tau-\tau') 
\nonumber \\
& & \times \exp [(eV\tau +i\phi (\tau))-(eV\tau' +i\phi (\tau'))]
\nonumber \\
& & [1+i e \eta(\tau)][1-i e \eta(\tau')]
\end{eqnarray}
describes the contribution of tunneling processes. In Eq.~(\ref{genfu3}),
$\alpha_{\rm ex}(\omega_l)$=$R_K |\omega_l|/4\pi Z_{\rm ex}(-i|\omega_l|)$
and $Z_{\rm ex}(\omega)$ is the Fourier coefficient of the external
impedance, where $\omega_l$=$2\pi l/\beta$ is the Matsubara frequency. For
a purely resistive impedance $R_{\rm ex}$, this expression is simplified
to $\alpha_{\rm ex}(\omega_l)$=$-\alpha_{\rm ex} |\omega_l|/4\pi$ with
$\alpha_{\rm ex}$=$R_K$/$R_{\rm ex}$. The $S_t[\phi,\eta]$ in
Eq.~(\ref{genfu4}) contains both the bias voltage and the driving source of
the generating functional. Here
$\alpha_t(\omega_l)$=-$\alpha_t|\omega_l|/4\pi$ with $\alpha_t$=$R_K$/$R_T$. 
It is important to notice that for $V$=0 and $\eta(\tau)$=$\eta(\tau')$=0,
Eq.~(\ref{zvpi}) reduces to the familiar form of the partition function of
an unbiased device\cite{l1}.  

Before analyzing the current, we would like to clarify what is indicated 
by the equations of the generating functional and the related actions. 
In derivation of the generating functional Eq.~(\ref{zvpi}), the technique 
that we used singles out the ground state, projecting out all excited states. 
At sufficiently low temperatures, our calculation should reproduce 
expectation values in the ground state. The ground state can be obtained from 
the non-equilibrium state at finite voltages by the transfer of a macroscopic 
number of electrons from one electrode to another. The exponential dependence 
on the voltage favors paths in which a large number of electrons move from one
electrode to the other at the beginning of the path, as implicit in
the action given by Eq.~(\ref{genfu4}).

The dc current $I(V)$ is readily derived as the first order functional
derivative of $Z_V[\eta]$ with respect to $\eta$,
\begin{eqnarray}\label{genfu5}
I(V) &=& 2e \int_{0}^{\beta} d\sigma \alpha_t(\sigma)
Z_V^{-1} \int D\phi e^{-S[\phi]} \nonumber \\
&\times& \, \frac{1}{2i} \left\{
\exp{(eV \sigma) \exp (i[\phi(\sigma)-\phi(0)])}
\right. \nonumber \\
& & \left. - \, \exp{(-eV \sigma)}
\exp{(-i[\phi(\sigma)-\phi(0)])} \right\} \, ,
\end{eqnarray}
where $Z_V$ is the path integral with the action
$S[\phi]$=$S_0[\phi]$+$S_{\rm ex}[\phi]$+$S_t[\phi,\eta$=$0]$. The current
auto-correlation function and high order correlation functions can be
obtained as well with the corresponding high order functional derivatives
of the generating functional $Z_V[\eta]$.  

Since in standard four-probe experiments, the tunneling current is measured
as a function of the average voltage drop $V_t$ across the tunnel junction,
instead of the bias voltage $V$, we should rewrite the above equation in a 
suitable form. This can be done by introducing the phase fluctuation of the
external impedance, $\theta(t)$=$\phi(t)$-$e(V-V_t)t$, into the path
integral expression for the tunneling current. Then, Eq.~(\ref{genfu5})
becomes
\begin{eqnarray} \label{cu} 
I &=& 2e \int_{0}^{\beta} d\sigma \alpha_t(\sigma)
Z^{-1} \int D\theta e^{-S[\theta]} \nonumber \\
&\times& \, \frac{1}{2i} \left\{
\exp{(eV_t \sigma) \exp (i[\theta(\sigma)-\theta(0)])}
\right. \nonumber \\
& & \left. - \, \exp{(-eV_t \sigma)}
\exp{(-i[\theta(\sigma)-\theta(0)])} \right\} \, ,
\end{eqnarray}
where $Z$ is the partition function.  The total action
$S[\theta]$=$S_c[\theta]$+$S_t[\theta]$+$S_{\rm ex}[\theta]$ in the above
equation is readily obtained from Eq.\ (\ref{sphi}) by setting $\eta$=0 and 
$V$=0, but with $\theta$ instead of $\phi$ as the variable.

Before going further, we would like to point out that the above general
expression reduces to the $P(E)$ theory\cite{dev90} at the weak tunneling
limit. To show this, we only need to neglect the contribution of the tunnel
junction to the total action, and then to perform an analytical
continuation from imaginary time to real time. Therefore, for tunnel
junctions with tunnel resistances much larger than the quantum resistance,
one can either use the $P(E)$ theory formulated in real time, or use our
Eq.~(\ref{cu}) represented in imaginary time.

In experiments, it is convenient to use Cr-films near the tunnel junction 
as well-controllable EME\cite{jed98,jyv98}. Consequently, from now on we 
will focus our attention on the case of purely resistive EME. Via a series 
expansion in $\beta eV$, we will derive the required linear and nonlinear 
response functions. We must point out that our results are not accurate if 
$\beta eV$ is large. Hence, when we compare our theory with experiments in 
the next section, we have checked that the theoretical results are valid 
under the conditions the experiments were performed.

\section{Results Compared with Experiments}

In this section, we will derive from Eq.~(\ref{cu}) some explicit results
suitable for comparing with experiments\cite{jyv98}. For this purpose we
will follow the approach of equivalent circuits, where the total action 
of the EME including the tunnel junction is replaced by the actions of an 
effective resistance and an effective capacitance. We will at first calculate 
the effective circuit parameters, and then use them to present the full I-V
characteristics of the tunnel junction.

We need to evaluate the partition function $Z$ in Eq.~(\ref{cu}). Let us first
write down the well-established partition function in the weak tunneling
limit
\begin{equation} \label{z0}
Z_0 = \prod_{l=1}^{\infty} (\beta\lambda^{(0)}_l)^{-1}
\end{equation}
where
\begin{equation}
\lambda^{(0)}_l = \frac{\omega_l^2}{2E_c}+
\frac{\alpha_{\rm ex}\omega_l}{2\pi}
\end{equation}
are eigenvalues with respect to the eigenfunctions of the action 
\[
S_0[\theta] \equiv S_c[\theta] + S_{\rm ex}[\theta] =
\sum_{l=1}^{\infty}\lambda^{(0)}_l(\theta_l'^2+\theta_l''^2) \, .
\]
Beyond the weak tunneling regime, the partition function, although more
complicated, can still be derived in the same manner. To the first order 
in $\alpha_t$, we have
\begin{equation}
Z = \prod_{n=1}^{\infty} \frac{ 1 + \int_{0}^{\beta} d\tau
\int_{0}^{\beta} d\tau' \alpha_t(\tau-\tau') \,
e^{f(\tau-\tau')} }{\beta \lambda^{(0)}_l} \, ,
\end{equation}
with the function $f(\tau)$ is defined as
\begin{equation}\label{cor}
f(\tau) = \frac{ \int D\theta \, \exp{ \{
-S_0[\theta(\tau)] \pm i\,[\,\theta(\tau)-\theta(0)\,] \}
}}{ \int D\theta \, \exp{ \{ -S_0[\theta(\tau)] \} }} \, .
\end{equation}

At sufficiently high temperatures, with a Taylor series expansion of
$\exp{[f(\tau-\tau')]}$, the partition function $Z$ can be expressed in the
same form as $Z_0$ in Eq.~(\ref{z0}), but with the eigenvalues 
\begin{equation} \label{eig}
\lambda_l = \frac{\omega_l^2}{2E_c^*} + \frac{\alpha^* \omega_l}{2\pi} \, .
\end{equation} 
Here the effective dimensionless conductance 
$\alpha^*$ = $\alpha_{\rm ex}$ + $\alpha_t$+$O(\beta E_c)$ contains the
contributions from both the external resistance and the tunnel junction. The
effective charging energy $E_c^*$, which is determined from
\begin{equation}\label{kac1}
\frac{1}{\beta E_c^*}=\frac{1}{\beta E_c} +
\frac{7.2\alpha_t\beta E_c}{8\pi^4}+O[(\beta E_c)^3] \, ,
\end{equation}
depends not only on the bare charging energy $E_c$, but also on the junction
conductance $\alpha_t$. The effect of the total EME is then embedded in the
effective capacitance $e^2/2E_c^*$ and the effective environmental
resistance $R_K/\alpha^*$. Consequently, from Eq.~(\ref{cu}) the transport 
properties can be investigated analytically\cite{man}, and the total
conductance
\begin{equation}
G(V_t) = G(0) + \delta G(V_t)
\end{equation}
has a voltage-independent term  
\begin{eqnarray} \label{li}
G(0) &=& \frac{1}{R_T} \{ 1-\frac{\beta E_c^*}{3}+
\left[ \frac{1}{15}+\frac{7.2\alpha^*}{4\pi^4} \right]
(\beta E_c^*)^2 \nonumber \\
&& +O[(\beta E_c^*)^3] \}
\end{eqnarray}
and a low-voltage correction
\begin{eqnarray} \label{nl}
\delta G(V_t) &=& \frac{1}{R_T} \{ 
\frac{ \beta E_c^* (\beta e V_t)^2}{45} + O[(\beta E_c^*)^3] \}
\nonumber \\
&& + O[(\beta e V_t)^4] \, .
\end{eqnarray}

The dots in Fig.~1 are the measured normalized conductance taken from the
Fig.~2a of Ref.~(\onlinecite{jyv98}). Using the sample parameter values
$\alpha_t$=1.04 and $\alpha_{\rm ex}$=8.09, as well as the experimental
temperature $T$=4.2~$K$, our calculated normalized conductance as a function
of the voltage is plotted in Fig.~1 as the solid curve. Our analytical
result is in very good agreement with the measurement not only for the
zero-voltage conductance, but also for the entire curve. We notice that the
experimental data exhibits conductance-step structures with sharp changes at
voltages $V_{\pm 1}$$\simeq\,$$\pm 0.2\,mV$ and
$V_{\pm 2}$$\simeq\,$$\pm 0.4\,mV$, a phenomenon which was not discussed in
the original experimental paper Ref.~(\onlinecite{jyv98}). Our conjecture is
that such conductance steps are originated from the resonances of the EME
modes. Using a single $LC$ mode as the EME and in the weak tunneling regime, 
Ingold and Nazarov\cite{gd92} have predicted the conductance step structures, 
which are sharp at low temperature but are smeared out at higher 
temperatures. If we estimate the inductance $L$ from the step width
$|eV_{\pm 1}|$=$\hbar /\sqrt{LC}$ with the capacitance in the range of $fF$,
we found $L$ in the range of $pH$, which is in agreement with the
experimental values\cite{jyv98}. These resonant EME modes are not included
in the purely Ohmic EME, but can be investigated if we use the
frequency-dependent external impedances to model the EME.

The temperature dependence of the zero-voltage conductance has been
measured thoroughly in Ref.~(\onlinecite{jyv98}). For the standard
four-terminal setup, if the temperature is not too low, the inverse of the
conductance dip at zero voltage  is linear in temperature
\begin{equation} \label{dg}
(\Delta G/G_T)^{-1}=G_T/[G_T-G(0)]=3 k_B T/E_c+\delta \, .
\end{equation}
Our theory gives the offset
\begin{equation} \label{del}
\delta=0.6+0.167 (\alpha_{\rm ex}+\alpha_t) \, , 
\end{equation}
while the $P(E)$ theory predicts
\begin{equation}
\delta_{P(E)}=0.6+0.167 \alpha_{\rm ex} \, . 
\end{equation}
Two sets of experimental data, taken from the Figs.~3a and 3b of
Ref.\ (\onlinecite{jyv98}), are plotted in Fig.~2 as dots, together with
the best fitted dotted lines. The sample parameter values are
$\alpha_t$=5.86, $\alpha_{\rm ex}$=20.32 and $C=$1.99~$fF$ for the upper set
of data, and $\alpha_t$=3.02, $\alpha_{\rm ex}$=1.50 and $C$=0.92~$fF$ for
the lower set of data. Using these parameter values, the analytical results
of our theory are plotted in Fig.~2 as solid lines, and those of the $P(E)$
theory are shown in dashed lines. While our theory agrees very well with the
measurements, the deviation of the $P(E)$ theory from the experiments
increases when the tunneling conductance gets larger. We have also performed
the self-consistent numerical calculations\cite{jed98} formulated in real
time, and the results are indistinguishable from our analytical solution
given by Eqs.~(\ref{dg}) and (\ref{del}).

\section{Analytical Solutions Based On Renormalized Parameters}

In the previous section we have shown that in the regime of not too low
temperatures, our strong tunneling formulas can be mapped onto the weak
tunneling formulas by renormalizing the charging energy and the external
impedance. Here we will analyze the situation of low temperatures. With
large $\beta$, the function $f(\tau)$ in Eq.~(\ref{cor}) can be written in
a simple analytical form
\begin{equation} \label{cor0}
f(\tau)= -\frac{2}{\alpha_{\rm ex}} \left( \gamma
+ \ln \frac{\alpha_{\rm ex} E_c |\tau|}{\pi} \right) \, ,
\end{equation}
where $\gamma$ is the Euler constant. With this expression we perform a
similar mathematical manipulation as what we have done for the case of high
temperature. For $\alpha_{\rm ex}$$\gg$1, the so-obtained eigenvalues have
again the same form as given by Eq.~(\ref{eig}), but with the effective
dimensionless conductance $\alpha^*$=$\alpha_{\rm ex}$+$O(1/\beta E_c)$,
and the effective charging energy
\begin{equation} \label{sel}
E_c^* = \left[ 1+\frac{\alpha_t 
(\pi\,e^{-\gamma}/\alpha_{\rm ex})^{2/\alpha_{\rm ex}}}{4\pi 
\Gamma (2/\alpha_{\rm ex})} \right]^{-1} E_c \, .
\end{equation}
In the above equation $\Gamma(x)$ is the Gamma function. At low temperatures
the renormalization is expected to be weak, because the tunneling processes
suffer strong Coulomb blockade. Using Eqs.~(\ref{cu}) and (\ref{cor0}), the
I-V curves are derived analytically as
\begin{equation} \label{pow}
I(V_t) = \frac{V_t}{R_T} \frac{1}{\Gamma (2+2/\alpha_{\rm ex})}
\left[\frac{\pi \exp (-\gamma) |eV_t|}{\alpha_{\rm ex} E_c^*} \right]
^{2/\alpha_{\rm ex}} \, .
\end{equation}

Now we have derived the effective conductance $\alpha^*$ and the effective
charging energy $E_c^*$ analytically for both low and high temperatures. Thus, 
we can map the nonperturbative expressions of the total action onto the 
perturbative results by replacing the bare parameters
$\alpha_{\rm ex}$ and $E_c$ with the effective parameters $\alpha^*$ and
$E^*_c$, respectively. Consequently, we can calculate the tunneling current
and the current-current correlation function beyond the perturbation theory.
A very important feature of our renormalization theory is the appearance of
the junction conductance in the effective charging energy $E_c^*$ as well as 
in the effective environmental conductance $\alpha^*$. Consequently, it
is inappropriate to  approximate the tunnel junction simply by an Ohmic 
element, although it is a good approximation at sufficiently high
temperatures\cite{l1,odi88,fle91,gwg}.

\section*{acknowledgment}
The authors would like to thank D.\ V.\ Averin, P.\ Joyez, 
A.\ N.\ Korotkov, J.\ P.\ Pekola, and J.\ J.\ Toppari  for valuable 
discussions. The work was support by the Swedish Natural Science Research 
Council under Grant Nos.\ F-FU 03996-312 and F-AA/FU 03996-313.

\appendix
\section*{}
In this Appendix we will derive Eq.~(\ref{genfu2}) for the path integral
representation of the generating functional of a voltage-biased tunnel
junction connected in series with an external impedance. Since the cases
we are interested in are not restricted to weak tunneling, we will use the
nonperturbative resummation technique.

>From Eqs.~(\ref{genfu}) and (\ref{htct}), we can expand the generating
functional in a Taylor series of the tunneling processes
\begin{eqnarray}
& & Z_V[\eta]= \sum_{l=0}^{\infty}(-1)^l
\int_0^{\beta} d\tau_l \int_0^{\tau_l} d\tau_{l-1} \cdots
\int_0^{\tau_2} d\tau_1 \nonumber \\
& & {\rm tr} \left\{ \exp [-\beta {\cal H}_0] {\cal H}_t(\tau_l) 
\cdots {\cal H}_t(\tau_1) \right\} \, ,
\end{eqnarray}
where 
\begin{eqnarray}
{\cal H}_0 &=& {\cal Q}^2/2C + \sum_{k,\sigma}
(\varepsilon_{k\sigma}+eV) c^{\dag}_{k\sigma} c_{k\sigma}
\nonumber \\
& & +\sum_{q,\sigma} \varepsilon_{q\sigma} c^{\dag}_{q\sigma} c_{q\sigma}
+\sum_{n=1}^{\infty} \left[ \frac{q_n^2}{2C_n} +
\frac{(\phi-\varphi_n)^2}{2e^2 L_n} \right] \, ,
\end{eqnarray}
and 
\begin{equation}
{\cal H}_t = \sum_{kq\sigma} \{ t_{kq\sigma}[1+ie\eta(\tau)]
c^{\dag}_{k\sigma}c_{q\sigma}\exp{(i\phi)}+{\rm H.c.} \} \, .
\end{equation}
Since the EME is modeled by a set of harmonic oscillators, the trace over
the  EME modes can be expressed explicitly as path integrals. By dividing
$\beta$ into $M$ segments of length 
$\varepsilon$=$\beta/(M$+$1)\rightarrow 0$ and then inserting eigenstates
of $\phi$ and ${\cal Q}$ between the elements of the Trotter product, we
have 
\end{multicols}
\noindent\hrulefill\hspace{0.5\textwidth}
\begin{eqnarray}
& & Z_V[\eta]={\cal N}' \sum_{l=0}^{\infty}(-1)^l \int_0^{\beta} d\tau_l 
\int_0^{\tau_l}
d\tau_{l-1} \cdots \int_0^{\tau_2} d\tau_1 
\prod_{m=1}^{\infty} \int D\varphi_m
\int d\phi_0 \int d\phi_1 \cdots \int d\phi_M \nonumber \\
& & \sum_{k_1^{}, q_1^{},
\sigma_1^{}, \xi_1^{}} \cdots \sum_{k_l^{}, q_l^{}, \sigma_l^{},
\xi_l^{}} 
\sum_{n_1^{}} \sum_{n_2^{}}\cdots \sum_{n_{M+1}^{}}
\exp \{ -\varepsilon [E_c n_1^2+C_m \dot{\varphi}_m^2/2e^2
+(\varphi_m-\phi_1)^2/2e^2L_m] \} \nonumber \\
& & \times \exp \{ in_1(\phi_0-\phi_1) \} 
\exp \{ -\varepsilon [E_c n_2^2+C_m \dot{\varphi}_m^2/2e^2
+(\varphi_m-\phi_2)^2/2e^2L_m] \} \exp \{ in_2(\phi_1-\phi_2) \} \cdots
\nonumber \\
& & \times \exp \{ -\varepsilon [E_c n_{M+1}^2+
C_m \dot{\varphi}_m^2/2e^2
+(\varphi_m-\phi_{M+1})^2/2e^2L_m] \} 
\exp \{ in_{M+1}(\phi_M-\phi_0) \} 
\nonumber \\
& &  
\times t^l \xi_1 \xi_2 \cdots \xi_l 
[1+ie\xi_l \eta(\tau_l)] \cdots [1+ie\xi_1 \eta(\tau_1)]
\prod_{\sigma} {\rm tr}_{\sigma} e^{-\beta
h_{\sigma}} h_{k_l^{}, q_l^{}, \sigma_l^{}, \xi_l^{}}(\tau_l) \cdots
h_{k_1^{}, q_1^{}, \sigma_1^{}, \xi_1^{}}(\tau_1) , \label{per}
\end{eqnarray}
\hspace{0.5\textwidth}\hrulefill
\begin{multicols}{2}
%
\noindent
where ${\rm tr}_{\sigma}$ is the trace over all electron states in the 
channel $\sigma$,
\begin{equation} \label{appenhs}
h_{\sigma}=\sum_{k} (\varepsilon_{k\sigma}+eV) 
c^+_{k\sigma} c_{k\sigma} +
\sum_{q} \varepsilon_{q\sigma} c^+_{q\sigma} c_{q\sigma}
\end{equation}
and 
\begin{equation} \label{appenhk}
h_{k,q,\sigma ,\xi}(\tau)= e^{\tau h_\sigma}
c^{\xi}_{k\sigma} c^{-\xi}_{q\sigma}
e^{-\tau  h_\sigma} e^{i \xi \phi(\tau)} \, .\label{hkq}
\end{equation}
To simplify our mathematical expressions, here we have introduced the
notations $c^{\xi}_{k\sigma}$=$c^+_{k\sigma}$ for $\xi$=1, and
$c^{\xi}_{k\sigma}$=$c_{k\sigma}$ for $\xi$=-1. The same notations apply
to $c^{\xi}_{q\sigma}$.

Using the Poisson's resummation formula to sum over $n_i$, we obtain
\begin{eqnarray}
& & \sum_{n_i=-\infty}^{\infty} e^{-\varepsilon E_cn_i^2-i
n_i\varepsilon \dot{\phi}_i}=\sqrt{\frac{\pi}{\varepsilon E_c}} 
\nonumber \\
& & \times \sum_{p=-\infty}^{\infty} e^{\pi p \dot{\phi}_i/ E_c
-\pi^2 p^2/\varepsilon E_c
-\varepsilon \dot{\phi}_i^2/4E_c} \, ,
\end{eqnarray}
where $\dot{\phi}_i$=$(\phi_i$-$\phi_{i-1})/\varepsilon$ as
$\varepsilon\rightarrow 0$. In the limit $\varepsilon E_c\rightarrow 0$,
only the $p$=0 term is relevant and all other terms are exponentially small.
Then we have
\begin{equation}
\lim_{\varepsilon E_c \rightarrow 0} \sum_{n_i=-\infty}^{\infty}
e^{-\varepsilon E_cn_i^2-i n_i\varepsilon \dot{\phi}_i}
= \sqrt{\frac{\pi}{\varepsilon E_c}} \,
e^{-\varepsilon \dot{\phi}_i^2/4E_c} \, .
\end{equation}
After the summations over $n_i$ have been performed,  the generating 
functional is simplified to the form
\end{multicols}
\noindent\hrulefill\hspace{0.5\textwidth}
\begin{eqnarray} \label{appen1}
& & Z_V[\eta]={\cal N}'' 
\sum_{l=0}^{\infty}(-1)^l \int_0^{\beta} d\tau_l \int_0^{\tau_l}
d\tau_{l-1} \cdots \int_0^{\tau_2} d\tau_1
\prod_{m=1}^{\infty} \int D\varphi_m 
\int d\phi_0 \prod_{i=1}^{M} \int d\phi_i 
\sum_{k_1^{}, q_1^{},
\sigma_1^{}, \xi_1^{}} \cdots \sum_{k_l^{}, q_l^{}, \sigma_l^{},
\xi_l^{}} 
\nonumber \\
& & \exp \{-\sum_{i=1}^{M} \varepsilon 
[\dot{\phi}_i^2/4E_c 
+C_m \dot{\varphi}_m^2/2e^2+(\varphi_m-\phi_i)^2/2e^2L_m] \} 
t^l \xi_1 \xi_2 \cdots \xi_l 
[1+ie\xi_l \eta(\tau_l)] \cdots [1+ie\xi_1 \eta(\tau_1)]
\nonumber \\
& & \times \prod_{\sigma} {\rm tr}_{\sigma} e^{-\beta
h_{\sigma}} h_{k_l^{}, q_l^{}, \sigma_l^{}, \xi_l^{}}(\tau_l) \cdots
h_{k_1^{}, q_1^{}, \sigma_1^{}, \xi_1^{}}(\tau_1) \, .
\end{eqnarray}
\hspace{0.5\textwidth}\hrulefill
\begin{multicols}{2}
%
\noindent

Now we need to evaluate the trace over the electronic states in the above
equation. This can be done by dividing the generating functional with a
constant term 
\[
\prod_{\sigma} {\rm tr}_\sigma \exp [-\beta (\sum_{k,\sigma}
\varepsilon_{k\sigma} c^{\dag}_{k\sigma} c_{k\sigma}+
\sum_{q,\sigma} \varepsilon_{q\sigma}
c^{\dag}_{q\sigma} c_{q\sigma})] \, ,
\]
which can be absorbed in an irrelevant prefactor ${\cal N}$. In this way
we have
\end{multicols}
\noindent\hrulefill\hspace{0.5\textwidth}
\begin{eqnarray}
& & Z_V[\eta]={\cal N} 
\sum_{l=0}^{\infty}(-1)^l
\int_0^{\beta} d\tau_l \int_0^{\tau_l}
d\tau_{l-1} \cdots \int_0^{\tau_2} d\tau_1 
\int D\phi \prod_{m=1}^{\infty} \int D\varphi_m 
\sum_{k_1^{}, q_1^{},
\sigma_1^{}, \xi_1^{}} \cdots \sum_{k_l^{}, q_l^{}, \sigma_l^{},
\xi_l^{}} 
\nonumber \\
& & t^l \xi_1 \xi_2 \cdots \xi_l 
[1+ie\xi_l \eta(\tau_l)] \cdots [1+ie\xi_1 \eta(\tau_1)] 
e^{-\int_{0}^{\beta} d\tau 
[\dot{\phi}^2/4E_c+C_m \dot{\varphi}_m^2/2e^2
+(\varphi_m-\phi(\tau))^2/2e^2L_m]} 
\nonumber \\
& & \times e^{\xi_l^{} [i \phi(\tau_l)+eV\tau_l]}
\cdots e^{\xi_1^{} [i \phi(\tau_1)+eV\tau_1]} 
\langle c^{\xi_l^{}}_{k_l^{} \sigma_l^{}}(\tau_l) 
c^{-\xi_l^{}}_{q_l^{} \sigma_l^{}}(\tau_l)
\cdots c^{\xi_1^{}}_{k_1^{} \sigma_1^{}}(\tau_1)  
c^{-\xi_1^{}}_{q_1^{} \sigma_1^{}}(\tau_1)\rangle _0 \, ,
\end{eqnarray}
\hspace{0.5\textwidth}\hrulefill
\begin{multicols}{2}
%
\noindent
where the symbol $\langle O \rangle_0$ denotes the quantum statistic average
over free quasi-particles,
\begin{equation}
\langle O \rangle_0=\frac{\prod_{\sigma} {\rm tr}_\sigma e^{-\beta
(\sum_k \varepsilon_{k\sigma} c^{\dag}_{k\sigma} c_{k\sigma}+
\sum_q \varepsilon_{q\sigma}
c^{\dag}_{q\sigma} c_{q\sigma})} O}{\prod_{\sigma} {\rm tr}_\sigma 
e^{-\beta (\sum_k \varepsilon_{k\sigma} c^{\dag}_{k\sigma} c_{k\sigma}+
\sum_q \varepsilon_{q\sigma} c^{\dag}_{q\sigma} c_{q\sigma})}} \, .
\end{equation}
Since only the combinations
$\langle c_{k_j^{}\sigma}^{\xi}(\tau)c_{k_i^{}\sigma}^{-\xi}(\tau')\rangle_0$
and
$\langle c_{q_j^{}\sigma}^{-\xi}(\tau)c_{q_i^{}\sigma}^{\xi}(\tau')\rangle_0$
are nonzero, we use the Wick's theorem to obtain 
\end{multicols}
\noindent\hrulefill\hspace{0.5\textwidth}
\begin{eqnarray} \label{cont}
& & \sum_{k_1^{}, q_1^{}, \sigma_1^{},  
\xi_1^{}} \cdots \sum_{k_l^{}, q_l^{}, \sigma_l^{}, \xi_l^{}} t^l 
\xi_1 \xi_2 \cdots \xi_l e^{\xi_l^{} [i \phi(\tau_l)+eV\tau_l]}
\cdots e^{\xi_1^{} [i \phi(\tau_1)+eV\tau_1]}
[1+ie\xi_l \eta(\tau_l)] \cdots 
[1+ie\xi_1 \eta(\tau_1)] 
\nonumber \\
& & \times \langle c^{\xi_l^{}}_{k_l^{} \sigma_l^{}}(\tau_l)
c^{\xi_l^{}}_{q_l^{} \sigma_l^{}}(\tau_l) 
\cdots c^{\xi_1^{}}_{k_1^{} \sigma_1^{}}(\tau_1) c^{\xi_1^{}}_{q_1^{}
\sigma_1^{}}(\tau_1) \rangle _0  \nonumber \\
&=& \sum_{pairs} \prod_{p=1}^{l/2} \sum_{\xi_p^{}}
\delta_{\xi_{p_1^{}}, -\xi_{p_2^{}}^{}}
\alpha_t(\tau_{p_1^{}}-\tau_{p_2^{}})
[1+ie\xi_{p_1^{}} \eta(\tau_{p_1^{}})] 
[1+ie\xi_{p_2^{}} \eta(\tau_{p_2^{}})]
\nonumber \\
& & \times  e^{\xi_{p_1^{}}^{} [i \phi(\tau_{p_1^{}})+eV\tau_{p_1^{}}]}
e^{\xi_{p_2^{}}^{} [i \phi(\tau_{p_2^{}})+eV\tau_{p_2^{}}]} \, ,
\end{eqnarray}
\hspace{0.5\textwidth}\hrulefill
\begin{multicols}{2}
%
\noindent
where
\begin{equation} \label{kern}
\alpha_t(\tau)=t^2 \sum_{k_1^{}, q_1^{}, k_2^{}, q_2^{}, \sigma} \langle
c_{k_2^{}
\sigma}^{\xi}(\tau) c_{k_1^{}\sigma}^{-\xi} \rangle_0 \langle c_{q_2^{}
\sigma}^{-\xi}(\tau) c_{q_1^{}\sigma}^{\xi} \rangle_0  \label{at} \, ,
\end{equation}
and the time-variables $\tau_{p_1^{}}$ and $\tau_{p_2^{}}$ are taken from the
values $\{ \tau_1, \tau_2, \cdots \tau_l \}$.
We notice that the contributions of the terms with four or more channel
indices equal are of order $1/N$. In metallic tunnel junctions, $N$ is very
large and so such terms can be neglected. Then for nonzero $\tau$, the
kernel function $\alpha_t(\tau)$ can be expressed as
$\alpha_t(\tau)$=$\alpha_t/4\beta^2\sin^2(\pi\tau/\beta)$, where
$\alpha_t$=$R_K/R_T$=$4\pi^2|t|^2\rho_l\rho_rN$ with $\rho_l$ (or $\rho_r$)
being the density of states in the left (or right) electrode. For Matsubara
frequencies $\omega_{\nu}$=$2\pi {\nu}/\beta$, the relevant Fourier
components reduce to the simple form 
$\alpha_t(\omega_{\nu})$=$-\alpha_t|\omega_{\nu}|/4\pi$. It is obvious that
the nonzero contributions to the sum over pairs in Eq.~(\ref{cont}) come
from those terms with $\sum_{i=1}^{l}\xi_i$=0 and even $l$. We let $l$=$2r$
be such even integers, and then because
\end{multicols}
\noindent\hrulefill\hspace{0.5\textwidth}
\begin{eqnarray}
& & \prod_{p=1}^{r} \sum_{\xi_p^{}}
\delta_{\xi_{p_1^{}}, -\xi_{p_2^{}}^{}}
\alpha_t(\tau_{p_1^{}}-\tau_{p_2^{}}) 
[1+ie\xi_{p_1^{}} \eta(\tau_{p_1^{}})] 
[1+ie\xi_{p_{p_2^{}}^{}} \eta(\tau_{p_2^{}})] \nonumber \\
& & \times e^{\xi_{p_1^{}}^{} [i \phi(\tau_{p_1^{}})+eV\tau_{p_1^{}}]
+\xi_{p_2^{}}^{}[i \phi(\tau_{p_2^{}})+eV\tau_{p_2^{}}]} \nonumber \\
& & =2^r \prod_{p=1}^{r}
\alpha_t(\tau_{p_1^{}}-\tau_{p_2^{}})
[1+ie \eta(\tau_{p_1^{}})] [1-ie \eta(\tau_{p_2^{}})]
e^{[i \phi(\tau_{p_1^{}})+eV\tau_{p_1^{}}]-
[i \phi(\tau_{p_2^{}})+eV\tau_{p_2^{}}]} \, ,
\end{eqnarray}
\hspace{0.5\textwidth}\hrulefill
\begin{multicols}{2}
%
\noindent
and because the sum over $r$ pairs gives a factor $(2r-1)!!$, we arrive at
the path integral representation of the generating functional 
\end{multicols}
\noindent\hrulefill\hspace{0.5\textwidth}
\begin{eqnarray} \label{appen2}
& &Z_V[\eta]=\prod_{m=1}^{\infty} \int D\varphi_m \int D\phi 
e^{-\int_0^\beta d \tau [\dot{\phi}^2/4E_c
+C_m \dot{\varphi}_m^2/2e^2+(\phi-\varphi_m)^2/2e^2L_m]}
\sum_{r=0}^{\infty} \frac{1}{r!} 
\nonumber \\
& & \times \left\{ \int_0^\beta d\tau
\int_0^\beta  d\tau' \alpha_t(\tau-\tau') 
e^{[(eV\tau +i\phi (\tau))-(eV\tau' +i\phi (\tau'))]} 
[1+i e \eta(\tau)][1-i e \eta(\tau')] \right\}^r \, .
\end{eqnarray}
\hspace{0.5\textwidth}\hrulefill
\begin{multicols}{2}
%
\noindent
This is Eq.~(\ref{genfu2}) in Sec. II.

\end{multicols}
 
%
 
\begin{figure}
\label{fig1}
\begin{center}
\includegraphics[angle=0,width=0.6\textwidth]{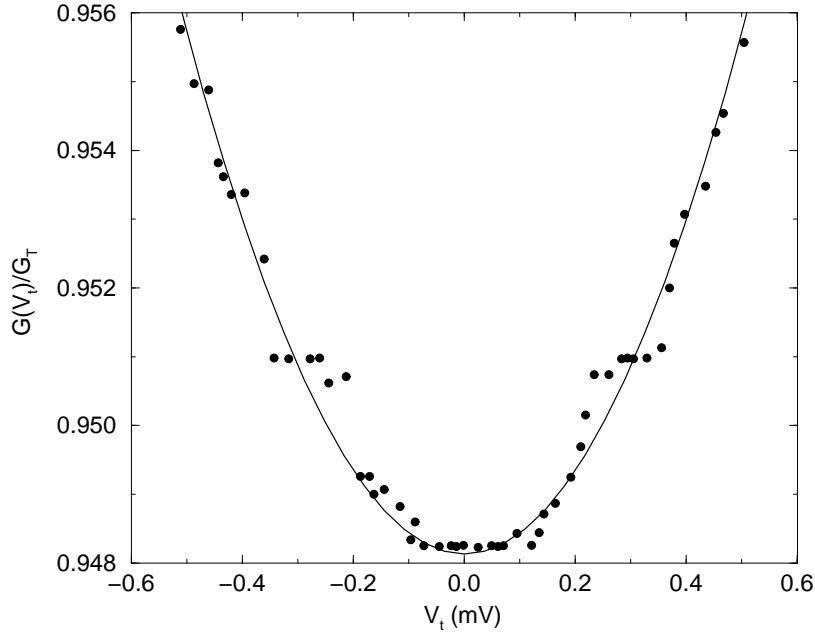}
\end{center}
\caption{Normalized conductance of the tunnel junction as a function of the 
junction voltage for $\alpha_t$=1.04, $\alpha_{\rm ex}$=8.09 and 
$\beta E_c$=0.18 at $T$=4.2~$K$.  The solid curve is from our theory, and 
the dots are experimental data taken from Fig.~2a of Ref.~(17).}
\end{figure}
\vfill

\newpage
\begin{figure}
\label{fig2}
\begin{center}
\includegraphics[angle=0,width=0.6\textwidth]{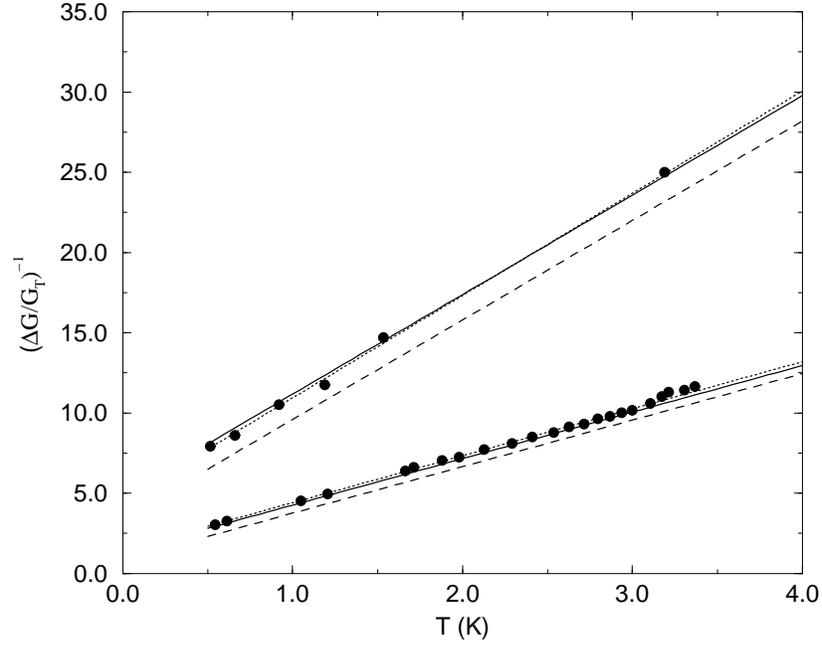}
\end{center}
\caption{Inverse of the normalized conductance dip at zero voltage as
a function of the temperature for $\alpha_t$=5.86, $\alpha_{\rm ex}$=20.32
and   $C$=1.99~$fF$ (upper curves); and $\alpha_t$=3.02,
$\alpha_{\rm ex}$=1.50 and $C$=0.92~$fF$ (lower curves). The dashed lines
are calculated from the P(E) theory, while the solid lines are calculated
from our present theory. The dots are the corresponding experimental data
taken from Figs.~3a and 3b of Ref.~(17), together with the best fitted
dotted lines.}
\end{figure}
\vfill

\end{document}